\documentclass[conference]{IEEEtran}
\ifCLASSINFOpdf
  \usepackage[pdftex]{graphicx}
  \graphicspath{{../pdf/}{../jpeg/}}
  \DeclareGraphicsExtensions{.pdf,.jpeg,.png}
\else
\fi
\begin{document}
\title{Joint Orchestration of Cloud-Based Microservices and Virtual Network Functions}

\author{\IEEEauthorblockN{Hadi Razzaghi Kouchaksaraei, Holger Karl}
\IEEEauthorblockA{Computer Network Group}
\IEEEauthorblockA{Paderborn University, Paderborn, Germany}
\IEEEauthorblockA{email: \{hadi.razzaghi, holger.karl\}@uni-paderborn.de}
}

%Telephone: (800) 555--1212, Fax: (888) 555--1212}


\maketitle


\begin{abstract}
Recent studies show the increasing popularity of distributed cloud applications, which are composed of multiple microservices. Besides their known benefits, microservice architecture also enables to mix and match cloud applications and Network Function Virtualization (NFV) services (service chains), which are composed of Virtual Network Functions (VNFs). Provisioning complex services containing VNFs and microservices in a combined NFV/cloud platform can enhance service quality and optimise cost. Such a platform can be based on the multi-cloud concept. However, current multi-cloud solutions do not support NFV requirements, making them inadequate to support complex services. In this paper, we investigate these challenges and propose a solution for jointly managing and orchestrating microservices and virtual network functions.

\end{abstract}

 \textit{\textbf{Keywords-Network Function Virtualization; Cloud Computing; Microservices; Virtualized Network Function.}}


% \fi
\IEEEpeerreviewmaketitle

\section{Introduction}
\label{sec:intro}
Today's cloud applications are commonly developed using a  microservice architecture. In this architecture, individual software components of an application are implemented as separate lightweight functional blocks, called microservices \cite{thones2015microservices}. This type of application can also smooth the road to the realisation of Distributed Cloud Computing (DCC) where microservices of a cloud application are deployed on geographically distributed micro data centres instead of on a single data centre.

Following the concept of cloud computing, Network Function Virtualization  (NFV) has emerged. It  aims at cloudifing network services that are conventionally provided by dedicated hardware. Similar to microservice-based or distributed cloud applications, NFV services consist of a set of distributed virtualised Network Functions (NFs) that are chained together to deliver a network service (e.g., Residential Gateway).

\begin{figure}[!ht]
    \centering
    \includegraphics[width=1\columnwidth]{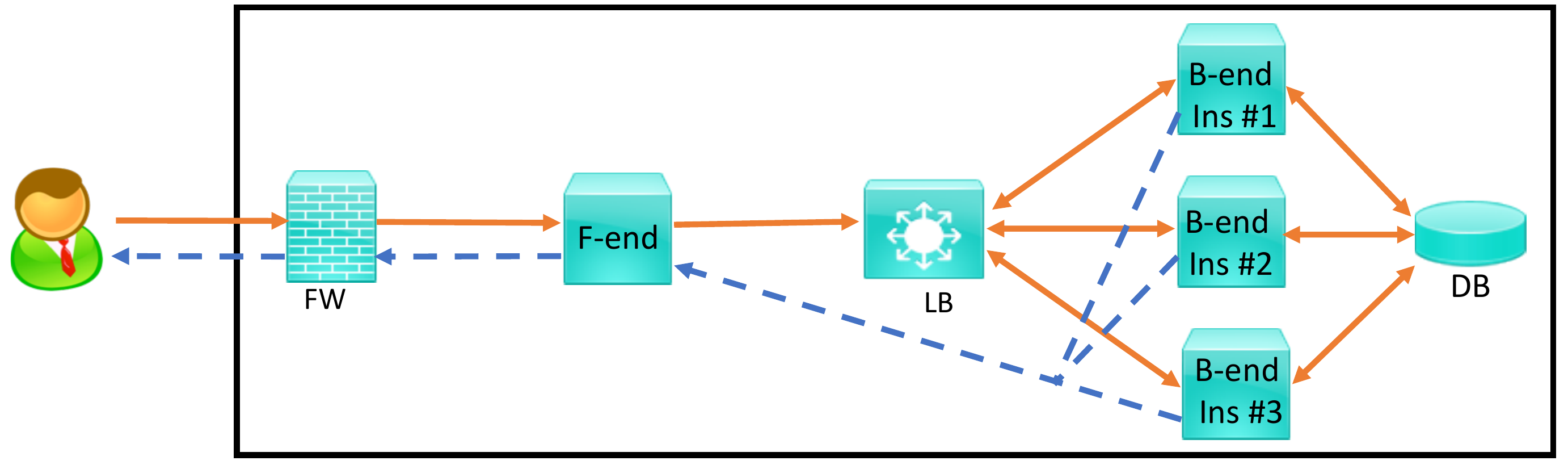}
    \caption{An example of complex services}
    \label{fig:complex}
\end{figure}

The distributed structure of NFV services and distributed cloud applications allows to mix and match VNFs and microservices. Such combined services, which we call \emph{complex services}, can have remarkable benefits for actors (users, service and infrastructure providers) involved in both NFV and cloud ecosystems such as cost optimisation and service quality improvement \cite{petcu2013multi}. An example of complex services (Fig.~\ref{fig:complex}) is a cloud application that includes a load balancer that spreads the load among application back-end instances and also a firewall that filters incoming requests to the application front-end. Since firewall and load balancer are network functions, they can have a better performance in NFV environments.

However, the fact that the current cloud and NFV platforms are not designed to support requirements of services from the opposite family makes the provisioning of complex services rather challenging. Some of the reasons for having different designs for cloud and NFV platforms are as follows:

\begin{itemize}

\item 
Packet processing is the main functionality of VNFs involved in NFV services, which should be performed as fast as possible using technologies such as Data Plane Development Kit (DPDK) \cite{ref:dpdk}. However, the majority of cloud applications do not need to bother with packet processing as it is not their primary objective.

\item 
Providing WAN connectivity is one of the essential tasks of an NFV
manager to provide service chaining for network services that are
geographically distributed. However,  in cloud ecosystems, providing
WAN connectivity is not crucial and is  therefore typically not offered by cloud management system such as OpenStack.

\item
The workload composition is also different in NFV and cloud ecosystems. While service chaining is used to combine VNFs in an NFV ecosystem, choreography/orchestration is used to connect microservices in a Cloud ecosystem.

\end{itemize}

A management and orchestration system that can join two ecosystems and manage all the differences can complement the shortcomings of cloud and NFV platforms in providing network functions and microservices, respectively. To this end, in this paper,  we propose a multi-cloud solution that unifies NFV and Cloud ecosystems. 

The paper is structured as follows. In Section~\ref{sec:cansolutions}, we shortlist the candidate solutions. In Section~\ref{sec:solutions}, the proposed solution is explained, and finally, in Section~\ref{sec:conclusion}, we highlight our conclusion.

\section{Candidate Solutions}
\label{sec:cansolutions}

An environment where NFV and cloud platforms can interwork with each
other could mitigate the challenge of provisioning complex
services. This could be realised by leveraging Multi-cloud
\cite{petcu2013multi} solutions, which allow VNFs and microservices to
be deployed on their respective platforms. Multi-cloud is an environment that uses the resources of multiple clouds (e.g., Amazon EC2, Windows Azure) to deploy a cloud application. Some of the goals of multi-cloud are to deal with peaks in service and resource requests, minimise cost, improve quality and availability, and  avoid cloud provider lock-in \cite{petcu2013multi}, which are similar to the objectives of providing complex services.

Although Multi-cloud can help the deployment of complex services in a combined NFV and cloud infrastructure, it cannot provide a comprehensive solution. Current multi-cloud solutions do not support NFV requirements, making them unsuitable for deployment and management of complex services. 

Terraform \cite{ref:terraform}, a multi-cloud solution that can handle
cross-cloud dependencies, supports most of the cloud management
systems such as OpenStack, K8, and AWS. However, Terraform does not
provide service chaining and WAN connectivity that is required for NFV
services. Cloudify~\cite{ref:cloudify} (another multi-cloud solution) is a composed NFV and cloud management and orchestration system. It supports the deployment of cloud and NFV services on multiple cloud infrastructures such as AWS and OpenStack. However, Cloudify does not allow the deployment of an application on multiple clouds at the same time~\cite{pham2015roboconf}, which makes it inadequate to manage and orchestrate complex services.

\section{Proposed Solution}
\label{sec:solutions}

\begin{figure}[!ht]
    \centering
    \includegraphics[width=1.0\columnwidth]{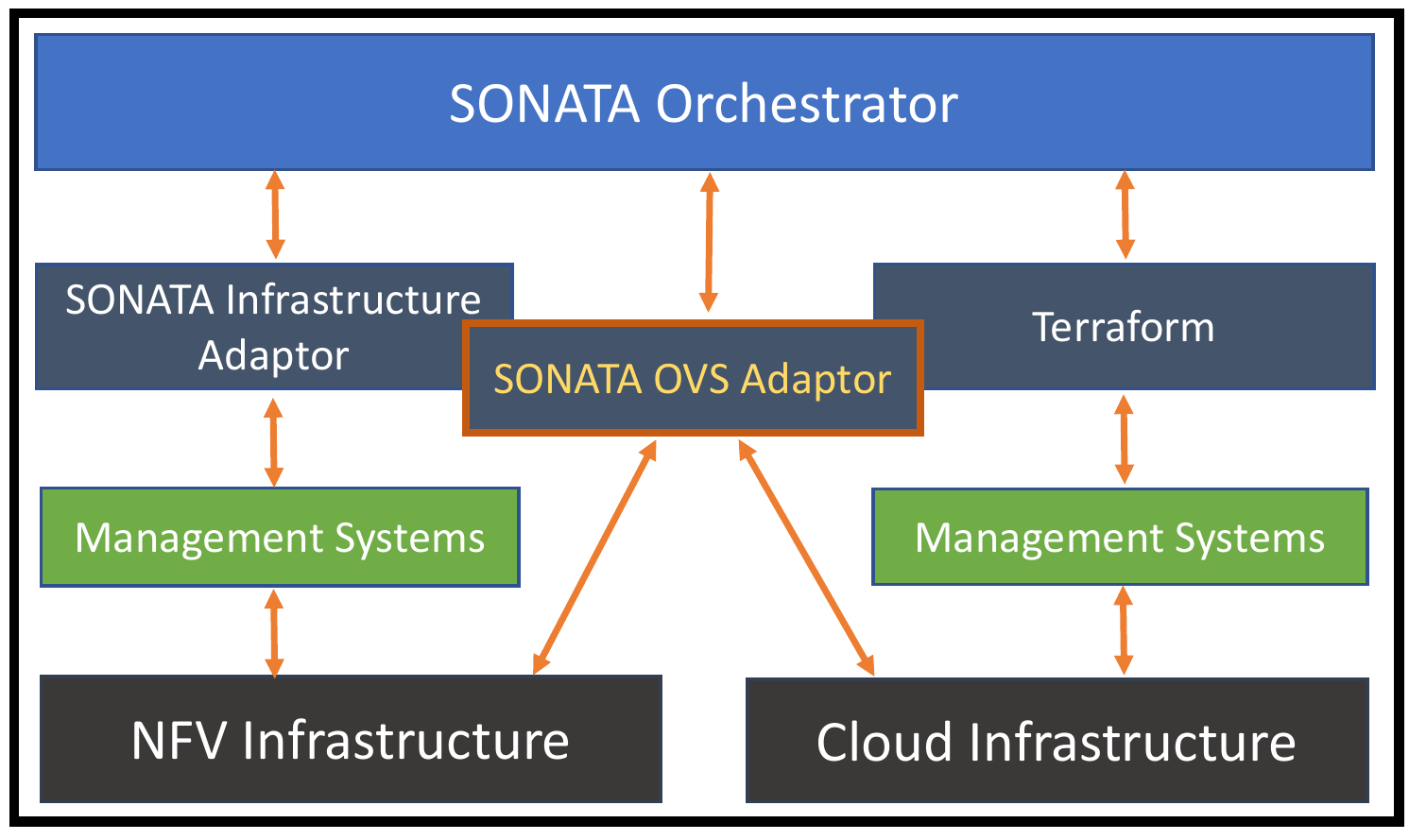}
    \caption{Proposed solution architecture}
    \label{fig:solution}
\end{figure}

Our solution is to consolidate current multi-cloud and NFV tools to
deploy, manage, and orchestrate complex services. To this end, in our
architecture (Fig.~\ref{fig:solution}), we combine
SONATA~\cite{ref:sonata}, a network service development and
orchestration platform, with Terraform. SONATA's orchestrator allows
services to be managed based on their specific requirements. This is a
valuable functionality for complex services which have management
requirements other than conventional network/cloud
services/applications. SONATA employs infrastructure and Open vSwitch
(OVS) adaptors to provide service chaining and WAN connectivity,
respectively. However, the infrastructure adaptor only supports
OpenStack and deploys services based on the NFV services
requirements. To solve this issue, we use Terraform to provide any cloud infrastructure for deploying microservices. Combining Terraform with SONATA infrastructure and OVS adaptors provides a unified NFV and cloud infrastructure that can be used by the orchestrator to deploy complex services.

Gathering the current tools and technologies for providing such an environment offers advantages such as reusability improvement and reducing maintenance overhead.

Our ongoing work includes the definition of a joint descriptor that can be used to describe both microservices and VNFs, as well as their dependencies, providing auto-scaling, and finding out the best way to distribute the life cycle management tasks between the orchestrator and management systems.

\section{Conclusion}
\label{sec:conclusion}
We investigated the overlooked area of managing and orchestrating complex services composed of VNFs and microservices. Reviewing the literature, we found that running complex services in a multi-NFV/cloud environment can have remarkable benefits for actors involved in both NFV and cloud ecosystems by reducing cost and improving the service quality. Our proposed solution for providing such an environment is to leverage tools and technologies that are used to realise multi-cloud environments. To this end, we are building a service platform that combines an NFV management and orchestration tool, SONATA, with a multi-cloud tool, Terraform. This combination provides a joint NFV and Cloud environment that can be used to deploy and manage complex services.

Our future work will be extending the proposed environment to support other cloud and NFV platforms to realise the deployment of complex services on any cloud infrastructure.

\section*{Acknowledgment}

\footnotesize{ This work has been partially supported by the SONATA project, funded by the European Commission under grant number 671517 through the Horizon 2020 and 5G-PPP programs and the 5G-PICTURE project, funded by the European Commission under grant number 762057 through the Horizon 2020 and 5G- PPP programs,}



\bibliographystyle{IEEEtran}
\bibliography{IEEEabrv,cloud}

\begin{thebibliography}{1}
\providecommand{\url}[1]{#1}
\csname url@samestyle\endcsname
\providecommand{\newblock}{\relax}
\providecommand{\bibinfo}[2]{#2}
\providecommand{\BIBentrySTDinterwordspacing}{\spaceskip=0pt\relax}
\providecommand{\BIBentryALTinterwordstretchfactor}{4}
\providecommand{\BIBentryALTinterwordspacing}{\spaceskip=\fontdimen2\font plus
\BIBentryALTinterwordstretchfactor\fontdimen3\font minus
  \fontdimen4\font\relax}
\providecommand{\BIBforeignlanguage}[2]{{%
\expandafter\ifx\csname l@#1\endcsname\relax
\typeout{** WARNING: IEEEtran.bst: No hyphenation pattern has been}%
\typeout{** loaded for the language `#1'. Using the pattern for}%
\typeout{** the default language instead.}%
\else
\language=\csname l@#1\endcsname
\fi
#2}}
\providecommand{\BIBdecl}{\relax}
\BIBdecl

\bibitem{thones2015microservices}
J.~Th{\"o}nes, ``{Microservices},'' \emph{IEEE Software}, vol.~32, no.~1, pp.
  116--116, 2015.

\bibitem{petcu2013multi}
D.~Petcu, ``{Multi-Cloud: Expectations and Current Approaches},'' in
  \emph{Proceedings of the international workshop on Multi-cloud applications
  and federated clouds}.\hskip 1em plus 0.5em minus 0.4em\relax ACM, 2013, pp.
  1--6.

\bibitem{ref:dpdk}
``\BIBforeignlanguage{english}{{DPDK}},'' {URL:} \url{http://dpdk.org/}
  [retrieved: January 2018].

\bibitem{ref:terraform}
``\BIBforeignlanguage{english}{{Terraform}},'' {URL:}
  \url{https://www.terraform.io/} [retrieved: January 2018].

\bibitem{ref:cloudify}
``\BIBforeignlanguage{english}{{Cloudify}},'' {URL:} \url{http://cloudify.co/}
  [retrieved: January 2018].

\bibitem{pham2015roboconf}
L.~M. Pham~et al., ``{Roboconf: a Hybrid Cloud Orchestrator to Deploy Complex
  Applications},'' in \emph{{IEEE 8th International Conference on Cloud
  Computing (CLOUD)}}.\hskip 1em plus 0.5em minus 0.4em\relax IEEE, 2015, pp.
  365--372.

\bibitem{ref:sonata}
S.~Dr{\"a}xler~et al., ``{SONATA: Service Programming and Orchestration for
  Virtualized Software Networks},'' in \emph{IEEE International Conference on
  Communications Workshops (ICC Workshops)}.\hskip 1em plus 0.5em minus
  0.4em\relax IEEE, 2017, pp. 973--978.

\end{thebibliography}

\end{document}